\documentclass[11pt]{article} 
\usepackage[utf8]{inputenc}

\usepackage{amsfonts}
\usepackage{amsmath}
\usepackage{amssymb}
\usepackage{amsthm}
\usepackage{caption}
\usepackage{color}
    \definecolor{darkgreen}{rgb}{0,0.5,0}
    \definecolor{darkblue}{rgb}{0,0,0.6}
    \definecolor{purple}{rgb}{0.4,.2,0.7}
\usepackage[margin = 2.5cm]{geometry}
\usepackage{graphicx}
\usepackage[hyperfootnotes = false, colorlinks = true, linkcolor = darkblue, citecolor = purple]{hyperref}
\usepackage{subcaption}
\usepackage{braket}
\usepackage{tikz}

\newcommand{\be}{\begin{equation}}
\newcommand{\ee}{\end{equation}}
\newcommand{\bea}{\begin{eqnarray}}
\newcommand{\eea}{\end{eqnarray}}
\def\la{\label}
\def\nref#1{(\ref{#1})}
\def\half{{1 \over 2 }}

\begin{document}

\thispagestyle{empty}
\begin{center}
    ~\vspace{5mm}
    
    {\LARGE \bf {String scale black holes at large $D$\\}}

    \vspace{0.5in}
    
    {\bf  Yiming Chen$^1$ and  Juan Maldacena$^2$}

   \vspace{0.5in}

    $~^1$Jadwin Hall, Princeton University,  Princeton, NJ 08540, USA \vskip1em
    
    $~^2$Institute for Advanced Study,  Princeton, NJ 08540, USA \vskip1em
     
    \vspace{0.5in}
    
%     {\tt   malda@ias.edu}

\end{center}

\vspace{0.5in}

\begin{abstract}
  
 We study aspects of black holes near the Hagedorn temperature.  The large dimension expansion introduced by Soda, Emparan, Grumiller and Tanabe connects them to the well studied two dimensional black hole based on $SL(2)_k/U(1)$. This allows us to explore black holes at string scale temperatures. We argue that the black hole can surpass the Hagedorn temperature, but at a somewhat larger temperature it is likely to turn over to a highly excited string. 
 
\end{abstract}

\vspace{1in}

\pagebreak

\setcounter{tocdepth}{3}

% {\hypersetup{linkcolor=black}\tableofcontents}

\tableofcontents

\section{Introduction}

\begin{figure}[t]
    \begin{center}
    \includegraphics[scale=.27]{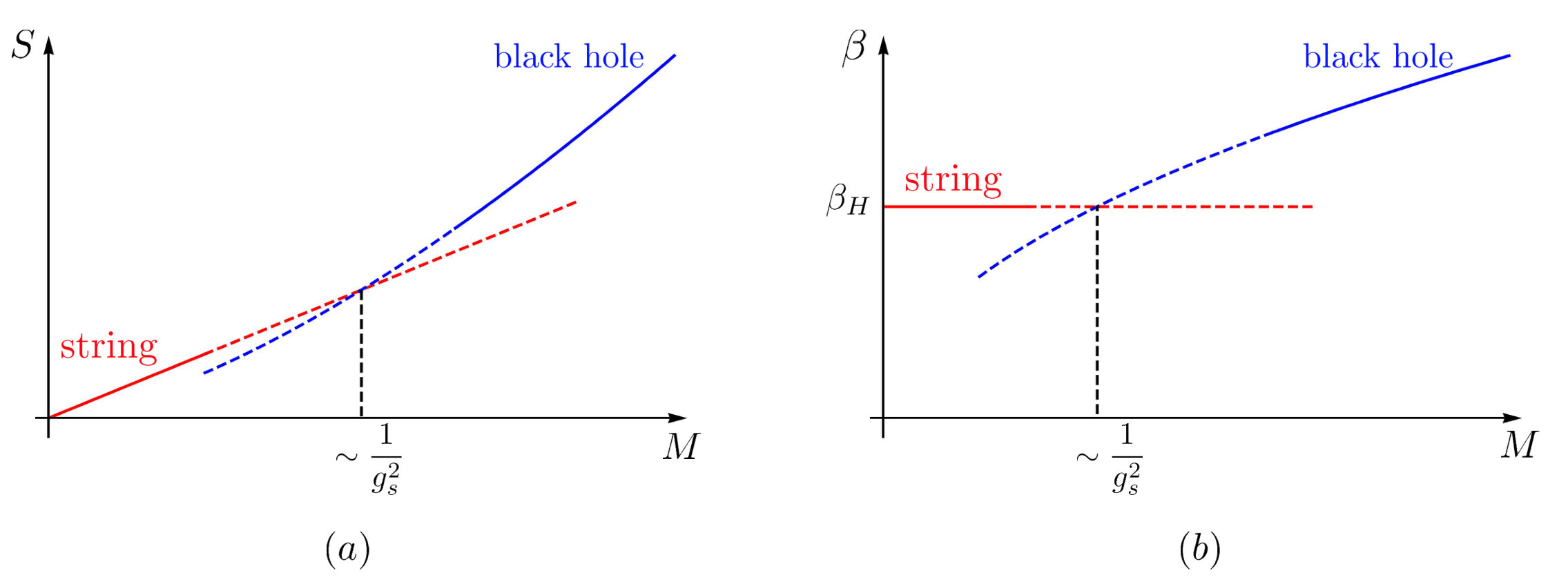}
    \end{center}
    \caption{(a) The entropy $S$ as a function of mass $M$ for both a free string and a black hole. The solid line describes the region where we can trust the computation and the dashed lines represent a naive (and incorrect) extrapolation. The extrapolated lines cross at roughly  $M \propto  1/g_s^2$ where we can trust neither of the solutions. At this point the black hole also has a string scale temperature, and a string scale size. }
    \label{EntropyMass}
\end{figure}

We consider black holes in a weakly coupled string theory, $g_s \ll 1$. They are well described by the usual Schwarzschild solution as long as their size is much larger than the string scale, $l_s = \sqrt{\alpha'}$. 
Interesting phenomena are supposed to occur when they approach this size. At this point their entropy is comparable to the entropy of an oscillating string of the same size, and it has been conjectured that they might transition into such oscillating strings \cite{Bowick:1985af,Horowitz:1996nw}. See figure \ref{EntropyMass}.  Understanding this transition in detail could be useful because on each side of the transition different aspects are manifest: on the string side, the microstates are manifest. On the black hole side, the interior is manifest.  With this general motivation in mind we will study some aspects of string size black holes. See \cite{Mertens:2015ola} for a nice review of this subject.

Let us first be a bit more precise on what is special about small black holes. The main issue is not so much the size but their Hawking temperature. In string theory there is a maximum temperature, called the Hagedorn temperature. As we approach it,  we can thermally excite more and more massive string states. The thermal ensemble is not even approximately defined beyond this temperature \cite{Hagedorn:1965st}.  
The thermal ensemble involves a thermal circle.  When this thermal circle has a critical length a winding string mode becomes massless. This mode describes several aspects of string thermodynamics. The theory near the Hagedorn temperature is described by an action of the form  \cite{Atick:1988si}
\bea \la{Act}
- I &=& { \beta \over 16 \pi G_N} \int d^{D-1}x \, \sqrt{g_D}  e^{ - 2 \Phi_D} \left[ {\cal R} + 4 (\nabla \Phi_D)^2 - |\nabla \chi|^2 - m^2 |\chi|^2 \right] 
\\ \notag &~& {\rm with}~~~~~~~~~~m^2 = {  \beta^2 - \beta_H^2 \over (2 \pi \alpha')^2 } ~,~~~~~~~~~ \beta^2 -\beta_H^2 \ll \alpha'~~
\eea  
where $\chi$ is the winding mode.\footnote{The winding mode is sometimes referred to as the ``winding tachyon", but it is not really tachyonic in most of what we will  consider.} The entropy of a configuration with non-zero $\chi$ is given by 
\be \la{EntWind}
S = { \beta \over 16 \pi G_N} \int  d^{D-1}x \, \sqrt{g_D} e^{ - 2 \Phi_D} { 2  \beta^2  \over (2 \pi \alpha')^2 } 
 |\chi|^2  
\ee 
This formula can be derived from  \nref{Act} as $S = (\beta \partial_\beta -1) I$, and taking the derivative only of the explicit $\beta$ dependence in the mass, since the $\beta$-derivatives of the rest of the fields vanishes by the equations of motions. A concrete solution where this formula applies is the self gravitating string solution  \cite{Horowitz:1997jc}. To derive \nref{EntWind} we assumed that the euclidean time circle never vanishes. 

For a Euclidean black hole the winding mode always has an expectation value, even at temperatures lower than the Hagedorn temperature. The reason is that winding is not conserved in the black hole background and therefore we can have an expectation value. More physically, the local temperature does become very large near the horizon and we can excite string states as we approach the horizon. So this winding mode could be qualitatively viewed as describing such strings \cite{Atick:1988si,Susskind:1994sm}.   
One interesting qualitative aspect of this winding condensate is that its  contribution to the entropy is formally of order $1/G_N \propto 1/g_s^2$, which has led some speculations that it could be the whole entropy of the black hole \cite{Susskind:1994sm}, though, to our knowledge, this has not been demonstrated. See \cite{Dabholkar:2001if} for ideas in this direction.  

\subsection{The two dimensional black hole background} 

A useful example of a black hole with possibly string scale curvature is the two dimensional black hole given by  
\be \la{Metr2d}
 ds^2 = k ( d\rho^2 + \tanh^2 \rho d\varphi^2 ) ~,~~~~~~~e^{ - 2 \Phi} = e^{ - 2 \Phi_0} \cosh^2 \rho  ~,~~~~~~\alpha'=1,
 \ee 
 where we wrote the Euclidean metric, valid for $k\gg 1$. From now on we will set $\alpha'=1$. 
 A nice feature of this solution is that there is an exact description based on a gauged WZW model, or 
 $SL(2)_k/U(1)$ \cite{Witten:1991yr,Dijkgraaf:1991ba}. With this description we can consider small values of $k$, which amount to string scale temperatures. 
 This allows one to see stringy phenomena which are not manifest in the metric description \nref{Metr2d}. The most interesting of these phenomena is the behavior of the condensate of the winding mode, which can be interpreted as the stringy atmosphere of the black hole. For large $k$ the winding condensate decreases very rapidly away from the horizon. As we lower $k$ it extends further and further away,  and at a critical value of $k$ it extends to infinity and becomes not normalizable \cite{Karczmarek:2004bw,Maldacena:2005hi,Kutasov:2005rr,Giveon:2005mi,Parnachev:2005qr}, making an infinite contribution to the   entropy of the black hole.   
 
One disadvantage of the two dimensional black hole is that the change of $k$, or the change of curvature radius translates into a change of the full theory, since $k$ is related to the total two dimensional central charge of the $SL(2)_k/U(1)$ theory. In other words, the black hole has an asymptotic geometry with a linear dilaton background whose physical gradient, $(\nabla \phi)^2 \propto { 1/ k}$, depends on $k$. 
 
\subsection{The large dimension limit}

In this paper, we start from the observation of \cite{Soda:1993xc,Emparan:2013xia}, who noticed that the large $D$ limit of a Schwarzschild black hole gives rise to the two dimensional black hole. 

We will first extend this observation to the stringy regime, to situations where the curvatures are of order string scale and argue that we can use the exact two dimensional black hole background to describe the system. 
As a check of the description we will recover the large $D$ limit of the $\alpha'$ corrections computed in 
\cite{Callan:1988hs}. We will argue that the black hole can have a temperature higher than the Hagedorn temperature due to the peculiar geometry near the horizon. However as we approach a slightly higher temperature the stringy thermal atmosphere\footnote{We can also call it the ``stringy corona".} expands and becomes a dominant contributor to the mass and entropy of the whole configuration. 

Using this string theory description we will also propose a computation of the chaos (or Lyapunov) exponent which takes into account all stringy corrections.

 \section{ Large $D$ black holes in string theory} 
 
 \subsection{Large $D$ black holes in gravity} 
 
 As shown in \cite{Soda:1993xc,Emparan:2013xia}, starting from the $D$ dimensional Schwarzschild metric 
 \be \la{EuSch}
 ds^2 = f dt^2 + { dr^2 \over f } + r^2 d\Omega_{D-2}^2 ~,~~~~~~~~f = 1 - {r_0^{D-3} \over r^{D-3} } ~,~~~~~~t \sim t+ \beta ~,~~~~~\frac{\beta}{2\pi} = {    2 r_0 \over D-3 }
 \ee 
 and defining 
 \be \la{RhoDef}
 \cosh^2 \rho = {r^{D-3} \over r^{D-3}_0} ~,~~~~~~~ \varphi = { 2\pi t \over \beta } 
 %~,~~~~~~ \frac{\beta}{2\pi} = {    2 r_0 \over D-3 } ,
 \ee 
 we get the metric \nref{Metr2d} with 
 \be \la{kGrav}
 k = \left( { 2 r_0 \over D-3} \right)^2 \sim \left( { 2 r _0 \over D } \right)^2~,~~~~~~{\rm for ~} ~~D\gg 1 ~,~~~~~\beta \gg 1 .
 \ee 
 Note also that the dilaton in \nref{Metr2d} is a two dimensional dilaton that can be viewed as arising from the volume of the sphere $e^{ -2\Phi} \propto r^{D-2} \sim \cosh^2 \rho $, for large $D$. 
 Note that the second inequality in \nref{kGrav} is necessary due to the fact that we have not yet included the stringy corrections. 
 Our goal is to extend this relationship to low values of $\beta$ where we have an exact stringy background. This will allow us to incorporate stringy corrections to Schwarzschild black holes and approach sting scale temperatures.

 \subsection{Large $D$ black holes in string theory} 
 
  \subsubsection{Large $D$ and the critical dimension} 
 
 The reader might question whether the large $D$ approximation is valid in string theory, where the dimension is fixed to be 10 or 26. One answer is that we can study first the worldsheet  CFT for large $D$, and then set $D=10$ or $D=26$ with the hope that the leading order in the large $D$ approximation gives us an accurate answer.   Whether this is the case or not depends on the precise form of higher order corrections in $1/D$, for which we provide some estimations in section \ref{sec:oneoverD}. 
 
  In addition, there is a context where the large $D$ {\it Euclidean} black hole can be embedded into string theory. We consider   a $(D+1)$ dimensional string theory with a timelike linear dilaton in the time direction and the Euclidean $D$ dimensional Schwarzschild background \nref{EuSch} in $D$ dimensions. Since the timelike linear dilaton can have arbitrarily negative central charge, we can then consider this as a full string background.\footnote{Notice that This works well in the bosonic string, or the type 0 string with $(1,1)$ worldsheet supersymmetry. For the type II superstring $D -2 =8 l  $ in order to have a standard  GSO projection \cite{Chamseddine:1991qu}.} Of course, if we already have a time direction, we cannot continue the Euclidean black hole into a Lorenztian one, since that would lead to a background with two time dimensions.

 \subsubsection{Constructing the worldsheet CFT at large $D$}
 
 As a first observation,  we notice from (\ref{RhoDef}) that, for large $D$, even with $\beta$ of order one, the value of $r_0$ would be of order $D$. Therefore the curvature radius of the sphere part is much larger than that of the two dimensional part involving the radial and time directions. This motivates us to treat the sphere directions perturbatively in $\alpha'$ but the two dimensional part exactly in $\alpha'$. 
 
Since the radius of the sphere is large (much larger than the string scale, $l_s$) we can approximate it as an almost CFT. This almost CFT has a central charge that is smaller than $D-2$ due to its curvature. 
In fact, we can use the formula for the central charge derived in \cite{Fradkin:1984pq} 
\be \la{cSphere}
c_{\rm{sphere}} = D-2 - { 6 \over 4 } {\cal R} ~\sim ~ (D-2) - 6 \left( { D \over 2 r } \right)^2 ~ \sim ~ D-2 - 6 \left( { D \over 2 r_0 } \right)^2 
\ee 
where ${\cal R}$ is the scalar curvature of the sphere, and we used the large $D$ approximation.  In principle, we could worry about higher order corrections to \nref{cSphere}. However, from large $D$ counting we expect that the answer should be of the form $ D f( D/r_0^2)$, so that given our scaling of $r_0$, $r_0 \propto D$,   the terms we have in \nref{cSphere} are all that we can get. Notice that when the proper distance in two dimensions varies over an order one amount (finite values of $\rho$ in \nref{RhoDef}), the value of $r$ barely changes away from $r_0$. Therefore, for finite values of $\rho$,  we can set $r\to r_0$ in \nref{cSphere}.
  This observation allows us to derive the   relation between $r_0$ and $k$, with $k$ defined as the level of the $SL(2)_k/U(1)$ coset theory. We need to use the exact expression of the two dimensional central charge of the $SL(2)_k/U(1)$ theory 
  \be \la{cTwo}
   c = 2 + { 6 \over k-2} .
   \ee 
  We then match the excess above $c=2$ in \nref{cTwo} with the deficit in \nref{cSphere} to obtain 
  \be \la{kex}
  { 6 \over k-2 } = 6 \left( { D \over 2 r_0 } \right)^2 ~~~~\longrightarrow ~~~~ 
  k-2 =  \left( {  2 r_0 \over D} \right)^2 .
\ee 
This corrects the gravity result \nref{kGrav}, and it goes over to it when $k$ is large. However, \nref{kex} is valid even when $k$ is of order one. 
In this discussion,  we are assuming that we are dealing with the bosonic string. We will discuss the type II superstring case in sec. \ref{sec:superstring}. 

In the two dimensional cigar CFT  describing the Euclidean black hole the conformal dimensions have the following form. They descend from formulas in $SL(2)_k$ and are parametrized by the $SL(2)$ spin $j$, and we have \cite{Dijkgraaf:1991ba}  (see also \cite{Giveon:2016dxe} for a more recent discussion and further references)  
\be \la{ExcCD}
\Delta = - { j(j-1) \over k-2} + {m^2 \over k } ~,~~~~~~\bar \Delta = - { j(j-1) \over k-2} + {\bar m^2 \over k } ~,~~~~~~~~~~~ m = \half ( n + k w) ~,~~~~~\bar m = \half (-n+ k w), 
\ee 
where $n$ is the momentum along the euclidean time circle and $w$ is the winding. Winding is not conserved in the background, but it is useful for labelling the states. When $j= \half + i s$, which is the continuous series representation of $SL(2)$, we can interpret $2s/\sqrt{k}$ as the momentum along the radial direction at infinity. We can read off the radius of the circle at infinity by comparing 
\nref{ExcCD} with what we expect for a linear dilaton background  to find that the radius of the circle at infinity is 
\be \la{Rex}
R = { \beta \over 2 \pi } = \sqrt{k}   = \sqrt{  \left( {  2 r_0 \over D} \right)^2 +2} 
\ee 
where we used \nref{kex} in the last equality. This gives us the \emph{full} stringy corrected relation between the radius $r_0$ and the temperature of the large $D$ black hole. The correction relative to the gravity answer in (\ref{RhoDef}) just comes from the replacement $k \to k-2$ we had in \nref{kex}. 

It would be highly desirable to have a formula for the entropy that we could apply to the cigar theory. Unfortunately, we do not have such a formula. 
 However, we expect that it should take the form 
 \be 
 S = e^{-2 \Phi_0 } F(k) ~,~~~~~~~~ e^{ - 2 \Phi_0 } \equiv { r_0^{D-2} \omega_{D-2} \over 4 G_N}  \la{EntG}
 \ee 
 where $\omega_{D-2}$ is the area of a $(D-2)$ unit sphere. $F$ is some function which would take care of the details of the cigar theory. $F$ should go to one for large $k$, which is the gravity limit. In this formula $r_0$  and $k$ depend on $\beta$ through \nref{Rex}. Using $ dM = T dS $ we can conclude that 
 \be 
 M =  { S \over \beta}  \left[ 1 + o (1/D) \right] \la{MassG}
 \ee 
 at large $D$. We get this formula by noticing that the main contribution to $dS$ comes from the term where we take the derivative with respect to $r_0$ in the term with $r_0^{D-2}$ in \nref{EntG}, since it leads to a factor of $D$. Notice that $S$ also depends on $\beta$. 
 
 We also conclude  that \nref{EntG} and \nref{MassG} acquire their main depedence on $\beta $ through the relation between $\beta$ and $r_0$ \nref{Rex}. In other words, we have 
 \be 
   \log S \sim D \log[ r_0(\beta) ] + o(D^0) = \log M + o(D^0) 
   \ee 
   This shows how to turn the stringy correction to $r_0$ in \nref{EntG} into more concrete physical observables. 

%\JM{I added the previous comment} 
 
\subsubsection{Checking against the first $\alpha'$ corrections}\label{sec:checkalpha}

We can check \nref{Rex} against the first $\alpha'$ correction which was computed in 
\cite{Callan:1988hs} for any $D$. They found\footnote{One subtlety is that the definition of $r_0$ in (\ref{TCMP}) from \cite{Callan:1988hs} is the radius of the sphere at the horizon in the \emph{Einstein} frame, not the \emph{string} frame that we work in. The difference between the two involves a factor $\exp\left(\frac{2\Phi_D}{D-2}\right)$ at the horizon, which becomes one in the large $D$ limit. Since we are not looking at $1/D$ corrections, we could ignore the difference and simply apply their formulas. }  
\be \la{TCMP}
{ \beta \over 2\pi } = \left({ 2 r_0 \over D-3 } \right) \left[ 1 + { (D-1) (D-4) \over 2 } { \lambda \over r_0^2 } \right]   ,
\ee 
where $\lambda =1/2 , ~1/4, ~0 $ for the bosonic,  heterotic  and type II strings respectively (recall we are setting $\alpha'=1$). 
We see that the large $D$ limit of  \nref{TCMP} reproduces the first non-trivial term in the $D/r_0$ expansion of  \nref{Rex}. 
 
 We can also look at the corrections to the metric and the dilaton. For the two dimensional black hole these were discussed in \cite{Dijkgraaf:1991ba}, see also \cite{Tseytlin:1991ht,Tseytlin:1993df,Jack:1992mk}. These can be found by adjusting the metric and the dilaton so that the wave equation reproduces the spectrum in \nref{ExcCD} (with $w=0$). In particular, for the dilaton one has 
 \be  \la{dilatonD}
 \Phi = \Phi_0- \log \cosh \rho - { 1 \over 2 ( k -2) \cosh^2 \rho } + \cdots ~,~~~~~~{\rm for }~~~~ k\gg 1
 \ee  
 We can view the term going like $\log\cosh \rho$ as arising from the dimensional reduction and the next term as arising from the $D$ dimensional dilaton.  In fact, this matches the large $D$ limit of the $D$ dimensional dilaton obtained at first order in $\alpha'$  in \cite{Callan:1988hs}\footnote{To take the large $D$ limit, it is convenient to start from the  expression (A.3) in \cite{Callan:1988hs} for $\phi'$ and then integrate it.}
 \be \la{dilatonD}
 \Phi_D = - \lambda \left( { D \over 2 r_0 }\right)^2 { r_0^{D-3} \over r^{D-3} }~,~~~~~~~~~\lambda_{\rm bosonic} = {\alpha' \over 2 } ~,~~~~\lambda_{\rm heterotic} = {\alpha' \over 4 } 
 \ee 
 See appendix   \ref{MetricApp} for further discussion. There we also match the $\alpha'$ corrections to the metric of the cigar CFT to the large $D$ limit of the corrections in \cite{Callan:1988hs}.

\subsection{The winding condensate for any finite temperature black hole}

In a black hole geometry,  the euclidean time circle shrinks to zero at the horizon. 
This means that winding along the circle is not conserved. In addition, it also means that winding modes can have expectation values \cite{Kazakov:2000pm,Kutasov:2005rr}. 

\begin{figure}[t]
    \begin{center}
    \includegraphics[scale=.15]{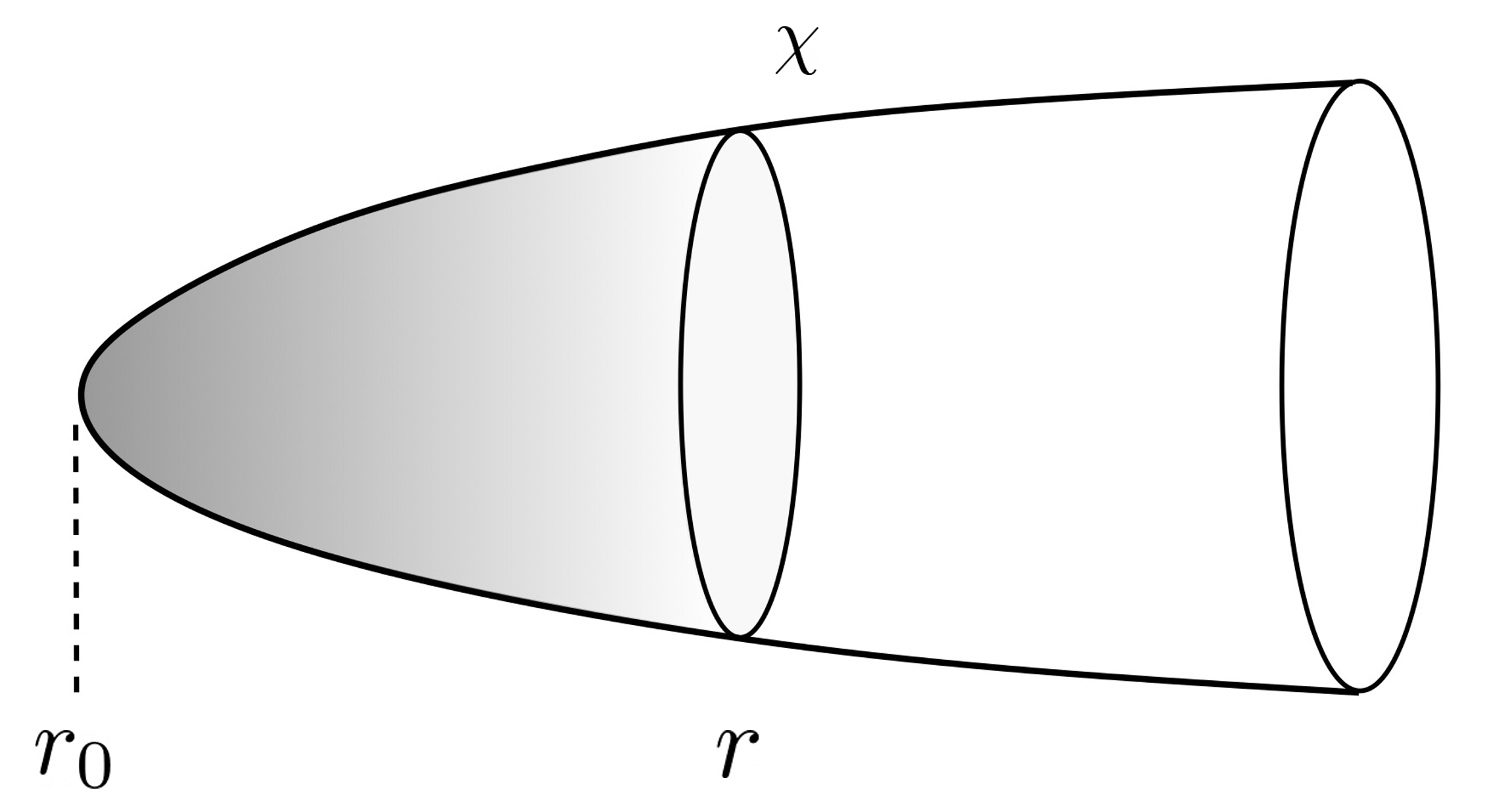}
    \end{center}
    \caption{We can estimate the expectation value of the winding condensate by considering a worldsheet that wraps the Euclidean black hole.}
    \label{WindingCondensate}
\end{figure}

Let us denote by $\chi$ the complex field associated to the winding mode. It is complex because the mode can wind in either directions of the circle. A simple estimate of its expectation value for any black hole is obtained by considering a worldsheet that wraps around the Euclidean black hole, see figure \ref{WindingCondensate}. This gives 
\be \la{chicond}
\hat \chi \propto  { 1 \over g_s} \exp\left( -{ \rm{Area}  \over 2\pi \alpha'}  \right) = { 1 \over g_s } \exp( -(r-r_0) R )  ~,~~~~~~~R = { \beta \over 2 \pi } \gg 1
\ee 
where $\hat \chi$ is the canonically normalized field, $\hat \chi = e^{ -\Phi} \chi$. 
The factor of $1/g_s$ is due to the fact that the topology of the worldsheet is a disk. This means that $\chi$ is of order one in the string coupling expansion, or a classical field. Of course, the exponential of the area gives a rapidly decreasing function. 
As $R$ gets smaller, we should include the ground state energy of the worldsheet fields on the cylinder and this replaces $R \to R - { R_H^2 \over 2  R } $ in \nref{chicond}. Here $R_H$ is defined as the radius corresponding to the Hagedorn temperature, $R_H = \beta_H /( 2\pi)$. This correction makes the winding condensate a bit larger. If we go to smaller values of $R$ then we need to include further fluctuations of the worldsheet and write a more proper expression 
\be 
 \chi \propto e^{ - m r } ~,~~~~~~~m^2 = R^2 - R_H^2 .
 \ee 
We recover the previous expressions by expanding in $R_H/R$. Here $m^2$ is the mass of the winding mode at large values of $r$. Here we are assuming that the dilaton is essentially constant.  
 
 We see that as $R \to R_H^+$, $m\to 0^+$ and the winding condensate becomes very large. 
 We can understand this as follows. In thermal equilibrium the black hole has an atmosphere of strings that start and end at the horizon 
 \cite{Susskind:1994sm}. As we approach the Hagedorn temperature this atmosphere is expected to grow and become larger and larger. Of course, this is not too surprising since already in flat space the thermal ensemble contains highly excited strings.\footnote{In fact, for the argument so far, we do not particularly care whether these strings are fundamental strings or weakly coupled strings of a large $N$ theory. In the latter case, we know what happens beyond $R_H$: we go to the deconfined phase.}    
  
The discussion so far  applies to any black hole at finite temperature.  

  \subsubsection{Winding condensate for the two dimensional black hole }\la{sec:winding2d} 
  
  For the particular case of the two dimensional black hole, we can also look at the winding mode. An important point is the following. Due to the dilaton gradient, there is a correction to the effective mass of the field, 
  \be \label{shiftmass}
  \int e^{ - 2 \Phi } |\nabla \chi |^2  ~~\to ~~~ \int |\nabla \hat \chi|^2 + (\nabla \Phi)^2 |\hat \chi|^2  ~,~~~~~~~~\hat \chi = e^{ -\Phi } \chi .
  \ee 
  We can interpret this as a correction to the Hagedorn temperature due to the decreased central charge of the rest of the dimensions. The   effective Hagedorn temperature is higher  since the string has less degrees of freedom to oscillate into. 
  \be 
  R^2_{H, \, \rm eff} = { c_{\rm transverse} \over 6 } ~,~~~~~~~c_{\rm transverse} =  { 24 - 6 (\nabla \phi)^2   }  ~,~~~~~~~~~R_H = { \beta_H \over 2 \pi } 
  \ee 
    This means that the total mass of the winding mode becomes 
  \be \label{TotalMass}
  m^2 = R^2 - R^2_{H, \, \rm eff} =  R^2 +  (\nabla \Phi)^2 - R_H^2 = { (k-3)^2 \over (k-2) } ~,~~~~~~R_H = 2 ~,~~~~~ \rho \gg 1
  \ee 
  for the bosonic string. 
  Interestingly, we see that it never becomes negative. It does becomes zero at $k=3$ and we will discuss the significance of that in a moment. 
    
  In the cigar theory, it is possible to calculate the asymptotic form of the winding condensate   \cite{Kazakov:2000pm} \footnote{This is sometimes called the Fateev-Zamolodchikov-Zamolochikov duality \cite{FZZ,Kazakov:2000pm,Fukuda:2001jd}.   However, we do not have to invoke this duality. The winding condensate is a feature that is present in both descriptions, the cigar and the sine liouville. See \cite{Hori:2001ax} for an argument for the supersymmetric version of the FZZ duality  and  \cite{Maldacena:2005hi} for the non-susy extension.}
       \be \la{WinPro}
  \hat \chi \propto  e^{ - ( k-3) \rho } 
  \ee 
  This is saying that the winding condensate is reasonable and normalizable for $k> 3$, but it becomes non-normalizable at $k=3$. 
  When it is very close to $k=3$ this winding mode makes a large contribution to the mass or the entropy of the system
  \be  \la{SWin}
  S_w \sim  \beta \int d^2 x\, \sqrt{g} \, 2 R^2 e^{ -2 \Phi} |\chi|^2  \propto  e^{ - 2 \Phi_0} \left( { 1 \over k-3}  + \cdots  \right)
  \ee 
 The dilaton dependence arises as we explained above. The factor of $1/(k-3)$ comes from integrating $|\hat \chi |^2 $ using 
 \nref{WinPro}. In particular we see that this contribution to the entropy of the system diverges as $k\to 3$. There is a similar expression for the mass, and it also diverges as $k\to 3$. 
 
The entropy \nref{SWin} contains only the winding mode contribution to the entropy, computed using the formula \nref{EntWind}. In principle there is also a black hole part which should also be of order $e^{ - 2\Phi_0}$, but without the factor of $1/(k-3)$. This means that as $k\to 3$ the winding contribution dominates the entropy. The full entropy of the system \nref{EntG} should include both corrections.

\subsubsection{Winding condensate for large $D$ black holes }\la{sec:windingD}

We have been discussing the two dimensional black hole, but this discussion can then be applied to the large $D$ black holes. We can match (\ref{TotalMass}) to the free equation of the winding mode in the large $D$ limit. The equation of motion for the winding mode $\chi$ in $D$ dimensions is
\begin{equation}\label{FreeWind}
	\frac{1}{r^{D-2}} \partial_r \left( r^{D-2}   \partial_r \chi  \right) -   \frac{\beta^2   - \beta_H^2}{(2\pi \alpha')^2 } \chi = 0 ~,~~~~~ { r^D \over r_0^D} \gg 1.
\end{equation}
We now focus on the region near but not too close to the horizon, or in the cigar coordinate $1\ll \rho\ll D$.   Using (\ref{RhoDef}) we write the equation  as 
\begin{equation}
   \frac{1}{k-2} \left[\partial_{\rho}^2 \chi + 2 \partial_\rho \chi \right] - \left(  k -4 \right)  \chi = 0 ~,~~~~~~{\rm or} ~~~~~~	 \frac{1}{k-2} \partial_{\rho}^2 \hat{\chi} - \frac{(k-3)^2}{k-2} \hat{\chi} = 0 ~,~~~~~~\rho \gg 1
\end{equation} 
with $\hat{\chi} = e^{-\Phi}\chi = \cosh \rho \chi$. 
%, then we have
%\begin{equation}
%	 \frac{1}{k-2} \partial_{\rho}^2 \hat{\chi} - \frac{(k-3)^2}{k-2} \hat{\chi} = 0.
%\end{equation}
Therefore we recovered the mass in (\ref{TotalMass}). In other words, the profile of the winding mode on the cigar matches the one near, but not too near,  the horizon of a large $D$ black hole. 
  
However, at large $D$, we should  also consider the winding mode profile far away from the horizon. The first observation is that when we reach the asymptotic Hagedorn temperature for the bosonic string, namely $R=R_H=2$, we have $k=4$. In the near horizon region, the one described by the cigar,  we can continue decreasing $k$, and increasing the local temperature,  without any problem  for $3<k < 4$. 
Of course, if we do this,  there will be a problem in the full geometry since the mass square of the winding mode will slowly decrease as we get away from the near horizon region and will eventually become negative. However, since $\chi$ is exponentially decreasing it will be very small when this happens. 

As the winding mode becomes massless at infinity, we expect that the integral of $|\chi|^2$ would produce a $1/m$ term in the mass and the entropy as $m\to 0$. This means we expect a $1/(k-4)$ pole in the mass or entropy of the black hole. However, this does not happen in the $D\rightarrow \infty$ limit. To see this, we can simply consider the special case of $k=4$, where the free massless equation (\ref{FreeWind}) gives two solutions $\chi \sim 1/r^{D-3}$ or $\chi \sim \textrm{constant}$. By matching with the profile $e^{-(k-2)\rho} = e^{-2\rho}$ on the cigar region, we see that we only have the $1/r^{D-3}$ part, and thus the winding mode is still normalizable. For finite $D$, we expect there can be a $1/(k-4)$ in the mass or entropy of the black hole. However, the coefficient of such a pole will be exponentially suppressed in $D$, since the winding mode has undergone a long period of exponential decay on the cigar region. 

Of course, if we are really considering the canonical ensemble, the fact that the winding mode becomes tachyonic at infinity for $3<k<4$ would be a problem.  However, we can imagine that we are in the microcanonical ensemble and we have a reasonable local thermal equilibrium near the horizon, but further away we are not in thermal equilibrium and we have some radiation coming out of the black hole. 

More precisely, if we have an evaporating black hole we expect that the temperature should be able to become higher than $T_H$, and still have a reasonably stable near horizon region. In other words, the black hole will produce a stringy halo which will be still localized near the horizon where the effective Hagedorn temperature is larger than the one at infinity, due to the reduced central charge of the sphere directions. 

However, as $k\to 3$, the stringy atmosphere expands and gets directly connected to the region where the winding mode becomes tachyonic and we do not get an approximate thermal equilibrium.

In principle we can only trust the description up to the point where $k-3 \sim 1/D$, which is the region where the black hole can be well approximated by the two dimensional cigar geometry. 
Everything that we said for the two dimensional black hole regarding the winding condensate and its contribution to the entropy applies here. 
In particular, we conclude that the winding condensate, or stringy atmosphere of the black hole, makes a larger fraction of the contribution to the entropy as $k\to 3$. 

One might be tempted to say that the entropy increases as $k\to 3$ due to the contribution of the winding condensate. However, in the large $D$ black hole the value of $e^{ - 2 \Phi_0}$, or the volume of the sphere, decreases rapidly as we decrease $r_0$. It decreases so rapidly that we cannot conclude that the total entropy increases as $k\to 3$, at least within the regime we trust our approximations. 

We expect that as an evaporating black hole reaches $k=3$ (or the corresponding temperature based on \nref{Rex}), it would  perhaps rapidly evaporate into a gas of strings. However, this is a dynamical process which we do not know how to treat.

\begin{figure}[t!]
\centering  

\begin{tikzpicture}[thick,scale = 1]

 \draw[double,->] (-1,0) -- (9,0);
 \draw (11.0,0) node{$k = \left(\frac{2r_0}{D}\right)^2 + 2 = {\beta^2 \over (2 \pi)^2} $};

\draw (-2.7,1) node{Cigar CFT};
\draw (-2.7,-1) node{Large $D$ black hole};

\draw (7,0) node[below]{$4$};
\draw (7.2, 0) -- (7.2,-1.5); 
\draw (9.2,-1.6) node[text width=3.5cm,text ragged] {\small Winding mode is massless at infinity ($\beta= \beta_H$)};

\draw (3,0) node[below]{$3$};
\draw (3.2, 1.5) -- (3.2,-1.5); 
\draw (5.2,1.6) node[text width=3.5cm,text ragged] {\small Winding mode becomes non-normalizable};
\draw (5.2,-1.6) node[text width=3.5cm,text ragged] {\small Winding mode becomes non-normalizable near horizon};

\draw (1,0) node[below]{$\frac{5}{2}$};
\draw (1.2, 1.5) -- (1.2, 0); 
\draw (1.2,1.6) node[text width=3.5cm,text ragged] {\small Lyapunov exponent vanishes};

\draw[dotted, ->] (3,-1) -- (1,-1) node[left]{?};

\end{tikzpicture}

\caption{We highlight some of the important values of $k$ or equivalently $\left( \frac{2r_0}{D}\right)^2 + 2$ where new qualitative features appear. The question mark indicates that the black hole picture stops to apply as the winding condensate becomes non-normalizable, and we don't know what the correct description of the full $D$ dimensional geometry is after that. The discussion on the Lyapunov exponent will be carried out in sec. \ref{sec:Lyapunov}.  }
\label{fig:kaxis}
\end{figure}
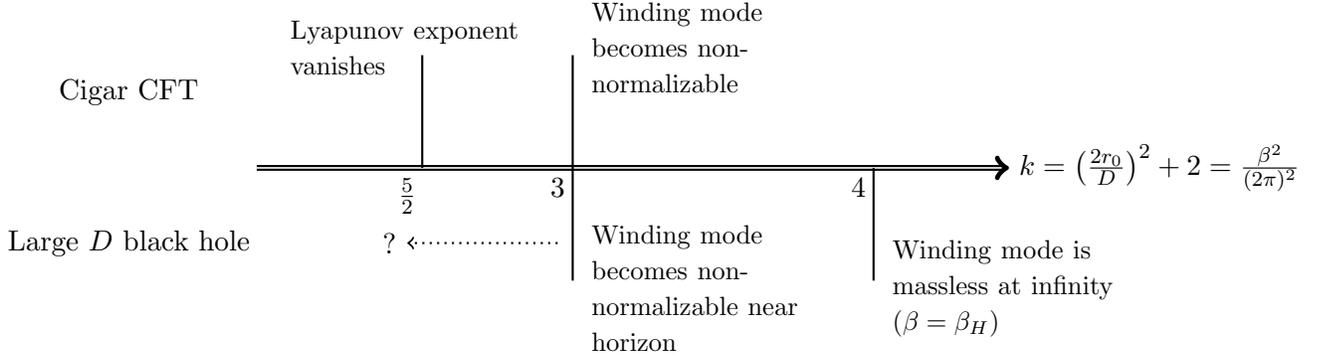

As a summary, in fig. \ref{fig:kaxis} we highlight some important values of $k$ or $\left( \frac{2r_0}{D}\right)^2 + 2$ in both the cigar CFT and the large $D$ black hole.

\subsection{The superstring case} \la{sec:superstring}

In this section we will discuss the superstring case, with $(1,1)$ worldsheet supersymmetry. We will not discuss the heterotic case. 
The supersymmetric version of the cigar theory was treated in \cite{Giveon:2003wn}. The level of the WZW model is shifted by two so that we replace $k-2 \to k$ in some of the formulas we had above. In particular, 
\be \label{susytemp}
k  = \left( { 2 r_0 \over D}\right)^2 ~,~~~~~~  {\beta \over 2 \pi } = R  = \sqrt{k}=   { 2 r_ 0 \over D } ~,~~~~~~~~~R_H  = \sqrt{2},
\ee 
where we also gave the value of the Hagedorn radius in flat space. 
 The fact that the temperature is uncorrected could be checked against the first non-zero $\alpha'$ correction, which is of order $\alpha'^3$, coming from the $R^4$ term. The correction is non-zero for general $D$. The formulas in \cite{Myers:1987qx} suggest that the correction is also non-vanishing at large $D$. However, some of the formulas there are incorrect and in a corrected analysis \cite{YCsusy}, the large $D$ limit of the correction indeed vanishes. 
 
%In this case, we do not expect any correction to the metric or the dilaton \cite{Jack:1992mk}. 

In this case, one might naively expect that there is no correction to the metric or the dilaton in the large $D$ limit. However, the correction to the dilaton is in fact non-vanishing, see \cite{YCsusy}.

The winding mode becomes non-normalizable as $k\to 1$, which is also smaller than $k=2$, which is the value that corresponds to the Hagedorn temperature at infinity. The physical interpretation is similar to the one of the bosonic string in sec. \ref{sec:windingD}.

 \subsection{Lyapunov exponent} \la{sec:Lyapunov}

 In this subsection, we use the exact CFT description of the cigar theory to propose a form 
 for the Lyapunov (or chaos) exponent \cite{Shenker:2013pqa,KitaevTalk} for the cigar theory.   In the range where the black hole picture is applicable ($k>3$), what we get is also the \emph{full} stringy correction to the Lyapunov exponent in the large $D$ limit.
 The idea is to use the exact CFT to compute the spin of the exchanged state in a scattering experiment that happens near the horizon. This is non-trivial because such states involve some analytic continuation. 
 For example, \cite{Brower:2006ea} considered the following vertex operator in flat space  and imposed the mass shell condition
 \be \la{PomFlat}
  V = ( \partial X^+ \bar \partial X^+)^{s/2} e^{ i q X} ~,~~~~~{\rm with}~~~~ 1 = { s \over 2 } + { \alpha' \over 4} q^2.
  \ee 
Analytically continuing in $s$ we can compute 
  $s = 2 +{\alpha' \over 2 } t$ with $t=-q^2$. Exchanging this operator produces the correct Regge behavior of scattering amplitude \cite{Brower:2006ea}.  
  
In the cigar CFT, in principle we can find the right state by considering the appropriate limit of a four point function. Instead, we will simply {\it guess} the right state. 
  
  In analogy with \nref{PomFlat} 
  we will look for a state with the form 
  \be \la{Can}
  |\Psi\rangle = ( J_{-1}^+ \bar J_{-1}^- )^{\gamma} |j,m,\bar m \rangle .
   \ee 
Here $J^a_n$ are the current algebra generators of the $SL(2)_k$ model, $j$ is the spin in $SL(2)$,  $m$ and $\bar m$ are the eigenvalues under $J_0^3$, $\bar J_0^3$. 
 This combination of currents is analogous to the ones in \nref{PomFlat} near the horizon, see e.g. \cite{Maldacena:2000hw}. One question is what values of $j$,  $m$ and $\bar m$ we should choose. These values are constrained by the $SL(2)/U(1)$ quotient to have the form 
 \be 
 \widetilde m =  m + \gamma = \half ( n + k w ) ~,~~~\widetilde {\bar m} = \bar m - \gamma = \half ( - n + k w ). 
 \ee 
 We will choose $w=0$. 
 
An important generator is the one performing the time translation, which is related to the rotation $B=J^3_0 - \bar J^3_0$, or the boost in the Lorentzian section. Our objective is to compute this boost eigenvalue $b$ since it then sets the Lyapunov exponent via 
 \be \la{lambdab}
 \lambda_L = { 2 \pi \over \beta } b .
 \ee  
We could arrange the scattering to be very close to the horizon by pushing the incoming particle towards early time. The state produced by the early incoming particle   has the property that it is very close to the past horizon. When the gravity approximation is good, the state will simply be a gravitational shock wave which travels along the horizon.  Away from the gravity limit, one way to characterize this in an algebraic way is the following. It is convenient to go to the covering $AdS_3$ problem where we have additional symmetry generators, such as $J^\pm_0$ $\bar J_0^\pm$. Under a boost that sends the particle to early times, some of these generators are sent to zero. Therefore, we expect that the candidate stringy version of the shock wave should be anihillated (or have very small eigenvalues) under such a generator.  
 This means that we want to impose 
 \be 
 J_0^+ |\Psi \rangle = \bar J^{-}_0 |\Psi \rangle \sim 0 
 \ee 
 This can be obeyed if we choose state in the continuous representation with $j =\half + i s $ and set $-m=\bar m =\half $. 
 The norm of $J^+_0 |j,m\rangle $ can be computed using the $SL(2)$ algebra via
 \bea \la{PairAct}
  J_0^- J_0^+ |j,m\rangle &=& \left[ \half ( J_0^- J_0^+ +  J_0^+ J_0^-) + \half[ J_0^-,J_0^+] \right] |j,m\rangle = 
  \cr &=&  \left[ 
  - j(j-1) + m^2 + m \right] |j,m\rangle 
  \eea 
 We see that it goes to zero as $j \to \half $ and $m= -\half$. A similar argument works for the state resulting from the action of $\bar J^-_0$. 
 Note that $J_0^+$ commutes with $J^+_n$ so that it goes through the first factor in 
 \nref{Can}. 
 Summarizing, 
 our candidate state is 
 \be 
 |\Psi \rangle = ( J_{-1}^+ \bar J_{-1}^- )^{\gamma} \left | j=\half + i s ,-\half ,\half  \right\rangle_{s \to 0 } 
 \ee 
 Imposing that it has   conformal dimension one,  we find 
 \be 
  1 = \gamma  - { j(j-1) \over k-2} + {  \widetilde m^2 \over k } = { b +1 \over 2 }  + { 1 \over 4 (k-2)} + { b^2 \over 4 k } ~,~~~~{\rm with} ~~~b = 2 \gamma -1 ~,~~~~~
 \widetilde m = \gamma -\half \ee 
  where $b$ is the total rotation (or boost) eigenvalue.  
 We then find that 
 \be \la{Exactb}
  b= - k + \sqrt{ k (k^2 -5) \over k-2} .
  \ee 
Plugging this into (\ref{lambdab}), we get the Lyapunov constant as a function of $k$.

 In the limit $k\gg 1$ we get that 
 \be 
 b = 1 - { 1 \over k } + \cdots ,
 \ee 
 which is in agreement with a direct gravity computation of the correction using the methods of \cite{Shenker:2014cwa} and the shock wave solutions in \cite{Sfetsos:1994xa}. 
 
 As $k$ increases $b $ goes to zero at $k = 5/2$. This value of $k$ is lower than three, so it is not useful for our black hole discussion, which was restricted to $k> 3$. Nevertheless, it is an interesting value of $k$. At this value of $k$ the central charge of the $SL(2)/U(1)$ theory is 14. This is the value that would appear in the superstring if we considered a purely two dimensional background which has total central charge 15 and the geometry of the cigar. The fermions would give a central charge of one, and the rest would give the central charge of 14. In analogy with \cite{Kazakov:2000pm} and using   the $\hat c=1$ matrix model discussion of \cite{Douglas:2003up}, we expect that this theory should have a dual description as a matrix model in non-trivial representations. In the singlet representation this model is integrable. So the fact that we get $b=0$ might be interpreted as evidence that also with non-trivial representations the model could be integrable. 
 
 We have not tried to understand what happens when $k< 5/2$.

\subsection{Estimation of $1/D$ corrections}\la{sec:oneoverD}

In this section we examine how well the large $D$ approximation works if we were to apply the results to $D=26$ for bosonic string, or $D=10$ for superstring. We will be using the relation between $\beta$ and $r_0$ as an example, which is   (\ref{Rex}) in the large $D$ limit. In this section only, we will define $r_0$ as the radius of the sphere at the horizon in the Einstein frame, but not the string frame, so that we can conveniently compare to the results in the literature. As we explained in a footnote in sec. \ref{sec:checkalpha}, the large $D$ limit yields the same relation between $\beta$ and $r_0$ regardless of which frame $r_0$ is defined in.

One issue is what the large $D$ expansion parameter really is. So far we've been loosely writing the expansion as one in $1/D$, but we could as well define it by shifting $D$ by an order one number. From our discussion, it is not obvious what the most natural choice is. One natural choice is to expand in $1/(D-2)$, since it is really the dimension of the $(D-2)-$sphere which we are taking to be large. Another choice would be to expand in $1/(D-3)$, since this is the combination appearing in the gravity expression (\ref{RhoDef}). In the following we use the latter choice. In other words, we write the large $D$ result (\ref{Rex}) for bosonic string as
\begin{equation}
	\frac{\beta}{2\pi} = \sqrt{ \left( \frac{2r_0}{D-3}\right)^2 + 2\alpha' } =\frac{2r_0}{D-3}  \left[  1 +  \frac{(D-3)^2 \alpha'}{4r_0^2}  + ... \right] = \frac{2r_0}{23}  \left[  1 +  \frac{529 \alpha'}{4r_0^2}  + ... \right]
\end{equation}
where we expanded to order $\alpha'$ and plugged in $D=26$. This should be compared with the finite $D$ answer in (\ref{TCMP}), which we copy below and plug in $D=26$
\begin{equation}\la{betad26}
	\frac{\beta}{2\pi} =\frac{2r_0}{D-3}  \left[  1 +  \frac{(D-1)(D-4) \alpha'}{4r_0^2}  + ... \right]=\frac{2r_0}{23}  \left[  1 +  \frac{550 \alpha'}{4r_0^2}  + ... \right].
\end{equation}
The relative error of the $\alpha'$ correction term is about $3.8\%$. Of course, one can complain that the reason the error is small here is because we chose $1/(D-3)$ as the expansion parameter, so that no error arises in the leading order result. If we instead expand in $1/(D-2)$, there will be error in both the leading term and the $\alpha'$ correction term, and the total error will be a function of $r_0$. We can consider the radius at which the winding mode becomes non-normalizable near the horizon, which is roughly at $2r_0/(D-3) = 2r_0/23 = \sqrt{\alpha' k} =\sqrt{3\alpha' }$. In this case, we find the error to be roughly $3\%$. Of course, it is not clear that we can trust (\ref{betad26}) at this small radius.

For the superstring case, the large $D$ result (\ref{susytemp}) is
\begin{equation}
		\frac{\beta}{2\pi} = \frac{2r_0}{D-3} = \frac{2r_0}{7},
\end{equation}
where we've plugged in $D=10$. The large $D$ answer does not contain any $\alpha'$ corrections. Therefore we can compare this with the finite $D$ answer containing the leading non-trivial $\alpha'$ correction \cite{YCsusy}:
\begin{equation}
\begin{aligned} \la{betasusy10}
		\frac{\beta}{2\pi} & = \frac{2r_0}{D-3} \left[ 1 +  \frac{\zeta(3)}{16}   \frac{(D-1)(4D^4 - 59 D^3 + 366 D^2 - 1113 D + 1350) \alpha'^3}{12 r_0^6}  + ...  \right] \\
		& \approx \frac{2r_0}{7} \left[ 1 +  \frac{440.6 \alpha'^3 }{ r_0^6}  + ...  \right]
\end{aligned} 
\end{equation}
where we've used $D=10$ in the second line. In this case, the relative size of the error depends on $r_0$, and is larger when $r_0$ is small. Similar to what we did above, we can look at the size of the error at the radius where the winding mode becomes non-normalizable near the horizon, which corresponds to $2r_0/(D-3) = \sqrt{\alpha' k } =\sqrt{\alpha'  }$. For $D=10$, the winding mode can become non-normalizable at some slightly different radius, but as a rough estimation, we'll simply take $2r_0/7 = \sqrt{\alpha'}$. Note that for this value it is not clear whether we should trust \nref{betasusy10}. Nevertheless,  we find that ratio between the correction term in (\ref{betasusy10}) and the leading term is about $24\%$. Instead, if we used $1/(D-2)$ as an expansion parameter, we would find an error of about $21\%$.

\section{Large $D$ limit of charged black holes}\la{sec:charge}

In this section, we show that the large $D$ limit of the near horizon geometry for a charged black hole matches with a two dimensional charged black hole background. Our discussion will be in the gravity limit, and we will leave a detailed comparison on the $\alpha'$ corrections to future study. 

We can start with a theory in $(D+1)$ dimensions and do a Kaluza-Klein reduction to $D$ dimensions on a circle. We consider configurations with both momentum and winding charge on the extra circle so that, on the worldsheet, we only couple to the left moving current of the extra circle. Of course, one could also consider general momentum and winding charges, but this case is a bit simpler because it leads to a solution where the extra circle has a constant size (in string frame).  The $D$ dimensional action   involves the metric, dilaton and a gauge field $A_\mu$
\begin{equation}
	I = \frac{1}{16\pi G_N} \int d^D x\, \sqrt{g_D} e^{-2\Phi_D} \left[ \mathcal{R} + 4 (\nabla \Phi_D)^2 - \frac{1}{4} F^2 \right].
\end{equation}
The charged black hole solutions of this action in general dimensions was presented in   \cite{Gibbons:1987ps} (see  
 \cite{Horowitz:1992jp}   for a nice review) 
\begin{equation}\label{metricchargeD}
\begin{aligned}
    ds^2 & = - \left(1 + \frac{r_0^{D-3} \sinh^2\alpha}{ r^{D-3}} \right)^{-2} f(r) dt^2 + \frac{dr^2}{f(r)} + r^2 d\Omega_{D-2}^2, \quad f(r) = 1 - \frac{r_0^{D-3}}{r^{D-3}},
\end{aligned}
\end{equation}
with gauge field and dilaton
\begin{equation}\label{gaugeD}
	A_{t} = \frac{ \sqrt{2} r_0^{D-3} \sinh \alpha \cosh \alpha}{r^{D-3} + r_0^{D-3} \sinh^2 \alpha},\quad e^{-2\Phi_D} = 1 + \frac{r_0^{D-3}}{r^{D-3}} \sinh^2 \alpha.
\end{equation}
$\alpha$ is a parameter of the solution, with $\alpha\rightarrow \infty$ being the extremal limit. Unlike the uncharged case ($\alpha=0$), the $D$ dimensional dilaton is excited. Also, note that we do not have an inner horizon, in contrast with the usual   Reissner-Norstrom solution.  We could do a Wick rotation on $t$ and take $\alpha\rightarrow i\alpha$ to go to the Euclidean signature. 

We would like to match the large $D$ limit of these solutions with 2d charged black holes \cite{McGuigan:1991qp}. The two dimensional charged black hole can be studied using the $\frac{SL(2)_k \times U(1) }{U(1)}$ coset model \cite{Johnson:1994jw,Giveon:2005jv}. Similar to what we had for the $D$ dimensional black hole, we consider a solution that is only charged under the left moving part of the extra $U(1)$. The background in the limit $k\gg 1$ is given by
\begin{equation}
	ds^2 = k \left[ d\rho^2 - \left( \frac{\tanh \rho}{1 - a^2 \tanh^2 \rho}\right)^2 d\varphi^2 \right],
\end{equation}
\begin{equation}\label{gauge2d}
	e^{ - 2 \Phi}  = e^{ - 2 \Phi_0}  \left( 1+ (1-a^2) \sinh^2 \rho\right), \quad A_{\varphi} = -\frac{\sqrt{2} a \sqrt{k} \tanh^2 \rho}{1- a^2 \tanh^2\rho}.
\end{equation}
$a$ is a parameter labelling the solution, with $a\rightarrow 1$ being the extremal limit.

We observe that under the identification
\begin{equation}
	 \cosh^2 \rho = \frac{r^{D-3}}{r_0^{D-3}} , \quad t \sim t+ \beta, \quad \frac{\beta}{2\pi}= \frac{2r_0}{(1-a^2) D}  ,\quad \sinh^2\alpha = \frac{a^2}{1-a^2},
\end{equation}
we get the metric in (\ref{metricchargeD}) with 
\begin{equation}
	k = \left(\frac{2r_0}{D} \right)^2,\quad \varphi = \frac{2\pi t}{\beta},\quad D\gg 1,\quad k\gg 1.
\end{equation} 
We also see that we have
\begin{equation}
	e^{-2\Phi_D} r^{D-2} = \left(1 + \frac{r_0^{D-3}}{r^{D-3}} \sinh^2 \alpha \right) r^{D-2} \propto 1 + (1-a^2) \sinh^2 \rho\propto e^{-2\Phi}
\end{equation}
in the large $D$ limit, which means that the two dimensional dilaton can be viewed as coming from the $D$ dimensional dilaton multiplied the volume of the sphere.

The gauge field in (\ref{gaugeD}) asymptotes to zero at infinity, while the one in (\ref{gauge2d}) goes to a constant value. This constant can be removed by a simple gauge transformation, after which the two expressions match.

   \section{Conclusions } 
   
   We have shown how the large $D$ limit introduced in \cite{Soda:1993xc,Emparan:2013xia} can be extended to the stringy regime with temperatures comparable to the Hagedorn temperature. This extension is relatively simple, all that needs to be understood is the proper map between parameters. This allows us to utilize the well studied $SL(2)_k/U(1)$ model,  but now reinterpreted as giving the near horizon geometry of the black hole. An interesting result is that these large $D$ black holes is that they  can have temperatures a bit higher than the Hagedorn temperature in the ambient space. Then we expect that an evaporating black hole would indeed get to these higher temperatures. 
   But we have not understood how it transitions into a gas of strings. 
   
   In the cigar CFT we can consider the model in the region $k< 3$  where the winding condensate is not normalizable. However, we have not managed to use it in any way that would shed some light on the full higher dimensional black hole.

    In the cigar theory, the FZZ duality \cite{FZZ} is often emphasized. It  connects the cigar to a theory on a cylinder with a winding condensate. This is interesting because the winding condensate could have a  Lorentzian descriptions with explicit microstates, see \cite{Giveon:2019gfk,Itzhaki:2019cgg,Jafferis:2021ywg} for  recent discussions. Presently,  this is poorly understood because there is a strong coupling region in the cylinder picture. In the Euclidean picture this is no problem because the string is prevented from going there by the sine Liouville potential. However, it is not clear how we are supposed to treat that region in the Lorentzian theory. 
   
       {\bf Acknowledgements }
  
    We thank L. Eberhardt, A. Goel, E. Witten and W. Zhao for discussions.
  
  JM was supported in part by U.S. Department of Energy grant DE-SC0009988 and by the It From Qubit Simons Collaboration.  
      
     \appendix 
     
\section{Matching the $\alpha'$ corrections to the metric at large $D$} 
\la{MetricApp}

In this appendix, we check that the $\alpha'$ corrections to the metric of the cigar match those of the large $D$ black hole. For the bosonic string, the leading correction to the Schwarzschild metric has the following form
\begin{equation}
	ds^2 = f \left( 1 + \alpha' \mu  \right) dt^2 + \frac{1}{f} \left( 1+  \alpha' \epsilon \right) dr^2 + r^2 d\Omega_{D-2}^2,
\end{equation}
where $f$ was given in (\ref{EuSch}). The functions $\epsilon,\mu$ are given in \cite{Callan:1988hs}, which in the large $D$ limit are\footnote{We work in the string frame as in other parts of the paper. It is useful to first apply the large $D$ limit to the differential equations for $\epsilon,\mu$ in Appendix A of \cite{Callan:1988hs} and then find the solutions.}
\begin{equation}\la{epsilonmu}
	\epsilon = \mathcal{O}\left( \frac{1}{D}\right), \quad \mu = - \frac{D^2}{2r_0^2} \frac{r_0^{D-3}}{r^{D-3}} +  \mathcal{O}\left( \frac{1}{D}\right).
\end{equation}

To compare this with the 2d metric, we need to first understand how the relation between $\rho$ and $r$ is corrected. This is determined by identifying
\begin{equation}
	{r^{D-2} \omega_{D-2} \over 4 G_N} e^{-2\Phi_D}   = e^{-2\Phi},
\end{equation}
where we are setting the asymptotic value of the $D$ dimensional dilaton to zero, $\Phi_D(\infty) =0$, by absorbing any such constant into $G_N$. 
The $D$ dimensional dilaton is given by \cite{Callan:1988hs}
\begin{equation}\la{correctionphiD}
	\Phi_D =  - \frac{\alpha'}{2} \left( \frac{D}{2r_0}\right)^2 \frac{r_0^{D-3}}{r^{D-3}},
\end{equation}
while the two dimensional dilaton is \cite{Dijkgraaf:1991ba}
\begin{equation}\la{correctionphi2d}
	\Phi = \Phi_0 - \log \cosh \rho - \frac{1}{2(k-2) \cosh^2 \rho} + ...
\end{equation}
where $\Phi_0$ is some constant. 
We see that the corrections in (\ref{correctionphiD}) and (\ref{correctionphi2d}) match with $(r/r_0)^{D-3} = \cosh^2\rho$. In other words,  the relation $(r/r_0)^{D-3} = \cosh^2\rho$ which was initially written down in the gravity limit in (\ref{RhoDef})  is actually unmodified at least to order $\alpha'$. Using this, we can then express (\ref{epsilonmu}) in terms of the 2d variables. We get
\begin{equation}
	\epsilon = 0, \quad \mu = - \frac{2}{k \cosh^2 \rho}.
\end{equation}
Since we are already looking at the leading correction terms, we do not need to distinguish $k$ and $k-2$ in the above formulas. 

For the $dt^2$ part of the metric we have
\begin{equation} \label{fdt2}
	f\left(1+  \alpha' \mu \right) dt^2 \approx   k \tanh^2 \rho \left( 1 - \frac{2}{k} \frac{1}{\cosh^2 \rho} \right) d\varphi^2,
\end{equation}
where we used the relation between $t$ and $\varphi$ is uncorrected. For the $dr^2$ part we have
\begin{equation}\label{fdr2}
	\frac{1}{f} (1+\alpha' \epsilon)  dr^2 \approx \frac{4r_0^2}{ D^2} d\rho^2 = (k-2) d\rho^2 ,
\end{equation}
where we used the corrected relation between $r_0$ and $k$ in (\ref{Rex}). 

Having expressed the leading $\alpha'$ corrections to the metric of the large $D$ black hole in the 2d variables, we can compare it with the known corrections in the cigar given in \cite{Dijkgraaf:1991ba,Tseytlin:1991ht,Tseytlin:1993df,Jack:1992mk}, which has the form
\begin{equation}
	ds^2 = (k-2) \left[ d\rho^2 + \frac{\tanh^2 \rho}{1 - \frac{2}{k} \tanh^2 \rho} d\varphi^2 \right].
\end{equation}
Expanding it to the leading order in $1/k$, we find a match with (\ref{fdt2}) and (\ref{fdr2}).

\section{Matching the gravity action with the cigar} 
\la{app:action}

In this appendix, we compute the Euclidean action of the Schwarzschild black hole, and discuss how it can be properly matched to the cigar. We will be neglecting all $\alpha'$ corrections, which we have not understood how to treat systematically. 

The action of the Schwarzschild black hole is given entirely by the Gibbons-Hawking-York boundary term \cite{GibbonsHawking}  
\begin{equation}\label{GHY}
	 I = - \frac{1}{8\pi G_{N}} \int d^{D-2}y \sqrt{h}\, (K - K_0)
\end{equation}
where we subtracted the extrinsic curvature for flat space $K_0 = \frac{D-2}{r}$. To evaluate the action, we need to cutoff the spacetime at a finite radius $r_c$. The standard discussion involves taking the limit $r_c\rightarrow \infty$, and we are left with a finite action
\begin{equation}\label{actionD}
	I = \frac{1}{D-3} \frac{\omega_{D-2} r_0^{D-2}}{4G_N}.
\end{equation}
We note that there is a $1/(D-3)$ factor in front. This factor is important when we use the action to compute the entropy
\begin{equation}
S = \left( \beta \partial_\beta - 1\right)I = \left( r_0 \partial_{r_0} - 1\right)\frac{1}{D-3} \frac{\omega_{D-2} r_0^{D-2}}{4G_N} =  \frac{\omega_{D-2} r_0^{D-2}}{4G_N}.
\end{equation} 
However, the factor $1/(D-3)$ in front seems unnatural if we view the action as coming from the cigar theory, since we would expect the action of the cigar to be proportional to $e^{-2\Phi_0}$, which is the second factor in (\ref{actionD}), without the extra $1/D$ factor. This puzzle can be resolved by putting the cutoff not at spatial infinity, but at the cigar region. We can define $(r/r_0)^{D-3} = \cosh^2 \rho$ as before, and put the cutoff at a finite value $\rho_c$. The action (\ref{GHY}) becomes
\begin{equation}\la{finitecutoff}
	I = \frac{1}{D-3} \frac{\omega_{D-2} r_0^{D-2}}{4G_N} \left[ 1 - (D-2) e^{-2\rho_c}\right].
\end{equation}
So far we have not done any approximation, and we can see that if we take $\rho_c \rightarrow \infty$, we recover (\ref{actionD}). However, now it is evident that if we put the cutoff at $e^{2\rho_c} \ll D$, in the large $D$ limit we will be left with
\begin{equation}\la{actioncigar}
	I = - \frac{\omega_{D-2} r_0^{D-2}}{4G_N} e^{-2\rho_c} = - e^{-2\Phi_0 -2\rho_c}.
\end{equation}
We see that the $1/D$ factor in front disappeared, and we have an expression that is consistent with the expectation from the cigar. An interesting point is that we need $e^{2\rho_c}\ll D$ to neglect the first piece in (\ref{finitecutoff}), but naively the cigar description can be applied as long as $\rho_c \ll D$. This suggests that the computation of the action is particularly sensitive to the cutoff. 

The cutoff dependence of (\ref{actioncigar}) agrees with a direct computation in the cigar theory \cite{Kazakov:2001pj}. In the cigar theory, the size of the circle is fixed at infinity. If we want to compute the variation of the action with respect to $\beta$, we would need to consider a finite cutoff at $\rho_c$, at which we have the size of the circle $\beta_c$, and vary $\rho_c$ such that $\beta_c$ varies by a small amount. At the same time, we should also adjust $\Phi_0$ so that the dilaton at the cutoff stays fixed.  

In other words, in the cigar theory we can fix the size of the circle and the value of the dilaton as  
\be 
 \beta_{c} = 2\pi \sqrt{k} \tanh \rho_c ~,~~~~~~~\Phi_c = \Phi_0 - \log \cosh \rho_c 
 \ee 
 The action evaluates to \nref{actioncigar}. As a check we can compute the entropy 
 \be 
 S = (\beta_c \partial_{\beta_c } -1) I = e^{ - 2 \Phi_0} 
 \ee 
 When we evaluate this derivative, it is important to keep $\Phi_c $ constant, which means that we need to vary both $\Phi_0$ and $\rho_c$ as we vary $\beta_c$. 

A key property of (\ref{actioncigar}) is that the action vanishes as we send the cutoff $\rho_c$ to infinity. A simple argument for this is that the cigar theory on its own has a zero mode $\Phi_0$, which will not be consistent with an $e^{-2\Phi_0}$ factor in the action unless the coefficient is zero. For this reason, we would expect to find zero from a string theory computation of the action, which makes it difficult to extract the thermodynamic quantities, especially the entropy.

\bibliographystyle{apsrev4-1long}
\bibliography{StringyBH}

\end{document}